\documentstyle[aps,epsfig,tighten,preprint]{revtex}
\begin{document}

\title{
Dependence of energy loss of jets on the initial thermodynamic
state of deconfined matter at RHIC
}

\author{
{\sc
K. Gallmeister$^a$, B. K\"ampfer$^a$, O.P. Pavlenko$^{a,b}$}
}

\address{
$^a$Forschungszentrum Rossendorf, PF 510119, 01314 Dresden, Germany \\[1mm]
$^b$Institute for Theoretical Physics, 252143 Kiev - 143, Ukraine
}

\maketitle

\begin{abstract}
The dependence of the radiative energy loss
of fast partons on the initial thermodynamic parameters is studied
for deconfined matter to be expected at RHIC.
We demonstrate that the specific QCD radiation pattern with a quadratic
dependence of the energy loss on the propagated distance leads to
a strong increase of the energy loss with increasing initial entropy
of deconfined matter supposed its life-time is less than the
average time to pass through the medium. 
This is in contrast to a parameterization
with constant energy loss per unit length of propagation.
For a sufficiently high initial temperature a two-regime
behavior of the energy loss as a function of the initial parton momentum
occurs.
The angular structure of the energy loss of hard jets with respect to the
initial temperature is also discussed for RHIC conditions.
\end{abstract}



\section{Introduction}

One of the primary goals of the current experiments at the
Relativistic Heavy-Ion Collider (RHIC) in Brookhaven National Laboratory
is the investigation of quark-gluon matter in a deconfined state.
While this new state of strongly interacting matter is thought to be
basically very different from a confined hadronic medium,
its verification in heavy-ion collisions seems to demand a combination
of various observables \cite{Heinz}.
Among the presently disputed signals of the quark-gluon plasma are
penetrating probes, like dileptons and photons, and hard probes,
like jets and the $J/\psi$, and accumulative probes,
like hadron multiplicity ratios. Since the hadron multiplicity measurements
reflect mainly the chemical equilibration and hadronization processes
in deconfined matter, they do not deliver direct information on the early
stage with maximum density and/or temperature, where a quark-gluon
plasma is expected to be formed. Dileptons and photons leave the
medium without strong interactions from the very beginning of the
matter's evolution and, therefore, can probe the initial stage of deconfined
matter. At the same time, the dileptons are emitted in all stages
so that the corresponding
spectra appear as a convolution of the production rate and
the whole space-time history of the matter. This leads to the possibility
to describe the secondary (thermal) dilepton and photon spectra,
at least for CERN-SPS energies, by one averaged temperature
\cite{NPA}. While the value of such an effective temperature is found
to be greater than the kinetic hadron freeze-out temperature, it still
gives only indirect information on the initial or hottest stage of deconfined
matter.

At RHIC, where one expects a much higher temperatures than at
CERN-SPS, the measurements of jets seem to offer an access to a probe
of the early stage  of the deconfined matter.
Being produced due to the hard initial collisions inside the volume
which is a moment later occupied by the quark-gluon medium, a
parent parton of a jet with high transverse momentum propagates
then for some time through the strongly interacting matter
and suffers an energy loss due to collisions and induced gluon radiation.
As a result, the momentum of the jet parton is attenuated before hadronizing.
This is the famous jet quenching anticipated in \cite{G_W}.
The energy loss in deconfined matter is estimated to be much larger than in
confined hadronic matter. Therefore, the energy loss is expected to
be most sensitive to both the size and the life-time
of the deconfinement stage.
One obvious way to study the system size dependence of jet quenching
is a variation of the centrality of the heavy-ion collision
\cite{LS,Wang_x}. Another opportunity is connected with variations
of the beam energy in central collisions. This enables one to investigate
the dependence of the energy loss on the life-time of the
deconfined system and, consequently, on the initial conditions, such
as temperature and parton density, expected to be achieved 
in RHIC experiments.

The dependence of the energy loss of a fast parton on the life-time and size
of the deconfinement region is also of particular interest due to the
non-trivial QCD behavior of the local radiative energy loss
with respect to the passed-through distance.
In QCD the radiative energy loss is mainly related to the coherence
conditions (Landau-Pomeranchuk-Migdal effect)
of the emitted gluons which suffer multiple non-Abelian
interactions within the ambient medium. In so far,
the radiative energy loss mechanism in QCD
has some analogy with the Ter-Mikaelian effect in QED
where the emitted photons suffer multiple Compton scatterings in the medium
(cf.\ \cite{TM} for
references and an explanation of these issues in the context of QCD).
As a result, a specific quadratic distance dependence of the total radiative
energy loss appears in deconfined matter.
This quadratic dependence is confirmed in various detailed calculations
\cite{Baier_1,Zakharov} and persists also in an expanding quark-gluon plasma
\cite{Baier_3}.
For further elaborate approaches to the energy loss we refer the
interested reader to \cite{G_L,Urs,Wang_33}
and for a recent application on the $p_\perp$ dependence
of pion and kaon ratios to \cite{jet_quenching}.

The aim of the present work is to study the dependence of the energy loss
on the initial thermodynamic variables of deconfined matter formed
in central heavy-ion collisions at RHIC. In particular, in section II
we search for an evidence of the specific QCD
pattern of the parton energy loss which reflects the appearance
of a high-temperature non-Abelian quark-gluon plasma.
These considerations are supplemented by estimates of the dependence
of the energy loss signal
on the jet opening cone (section III). The discussion and conclusions
can be found in section IV.

\section{Energy loss effects of fast partons}

According to \cite{Baier_3} the total radiative energy loss of a parton
propagating through an expanding quark-gluon plasma is given by
$\Delta E = \xi \Delta E \vert_{T_f}^{\rm static}$,
where $\xi = 2$ (6) for a parton created inside (outside) the medium.
The medium is assumed to expand and obeys the temperature evolution
$T \propto \tau^{-\beta}$ with $\tau$ as proper time in the piece
of matter undergoing boost-invariant longitudinal expansion,
and $\beta \le 1$. $T_f$ denotes the plasma temperature at which the
parton leaves the deconfined region either through a time like
or a space like boundary.
$\Delta E \vert_{T}^{\rm static}$ corresponds to the energy loss in
a static system with temperature $T$ and is given by \cite{Baier_1}
\begin{equation}
\Delta E \vert_{T}^{\rm static} = - \frac 14 \alpha_s N_c \times
\left\{
\begin{array}{ll}
\hat q L^2, & L < L_{\rm crit}\\
\sqrt{\hat q E} L, & L > L_{\rm crit}
\end{array}
\right. ,
\end{equation}
where $L$ is the distance in transverse direction the fast jet quark
propagates till escaping the region with deconfined matter;
$\hat q$ denotes a transport coefficient, and $N_c = 3$.
The critical length $L_{\rm crit}$ depends on the initial quark energy as
\begin{equation}
L_{\rm crit} = \sqrt{\frac{E}{\hat q}}.
\end{equation}
The transport coefficient is approximated in the subsequent calculations
by $\hat q = \mu^2 / \lambda$ with $\lambda$ as mean free path of the
partons and $\mu$ as the average transverse momentum kick which the
fast quark suffers per scattering while propagating through the deconfined
medium. We use the screening mass as scale for $\mu$.
The temperature dependence of the parameters describing the medium
is as follows: $\lambda^{-1} = 2.2 \alpha_s T$ \cite{Biro},
$\mu^2 = 6 \pi \alpha_s T^2$ \cite{Levai}, and $\alpha_s = 0.3$.
We use here $\xi = 2$, i.e., the lower limit of the QCD energy loss
from \cite{Baier_1}.

One should keep in mind that the estimates of the energy loss based
on the above described equations are semi-quantitative due to the following
reasons.
(i) The parameter $\hat q$ is model dependent. Accordingly the numerical
value of the energy loss according to Eq.~(1) can change by a constant
factor of about 2.
(ii) One can also expect an additional suppression factor in Eq.~(1)
if the number of parton scatterings inside deconfined matter is not
large and one has to use an opacity expansion \cite{G_L}
(cf.\ also \cite{Schaefer}).
However, the most important $L^2$ dependence of the
energy loss and the specific QCD pattern governed by the
parameter $L_{\rm crit}$ are expected to be stable with respect to
variations of the medium dependent parameters.
(For a recent discussion of the $L^2$ and $L_{\rm crit}$
dependences cf.\ \cite{Urs}.)
(iii) Equations (1, 2) are based on one-gluon emission. A reliable
estimate of multi-gluon emission is not yet available. Nevertheless,
we use Eqs.~(1, 2) as benchmark and explore in the following their
consequences.

To simplify our considerations we model the space-time evolution of the
quark-gluon plasma by the Bjorken scenario \cite{Bjorken}
with boost-invariant longitudinal expansion and conserved entropy per
rapidity unit. This results in the life-time of deconfined matter
$\tau_c = (T_i/T_c)^3 \tau_i$, where $T_i = T(\tau_i)$ is the ''initial''
temperature at ''initial'' time $\tau_i$.
(Actually, $T_i$ and $\tau_i$ refer to the earliest stage where
the partonic system is sufficiently near to equilibrium.)
$T_c \approx 170$ MeV
is the temperature where in a baryon-poor plasma the confinement sets
in according to QCD lattice calculations \cite{Karsch}.
We also assume that transverse expansion is negligible for the early
(deconfined) state of matter.

Taking Eq.~(2) as a dynamical ingredient we calculate with Monte Carlo
simulations the average energy loss of a fast quark created randomly,
but with uniform distribution over the central slice of matter
at mid-rapidity, with random transverse
direction. Due to the fast longitudinal expansion of the matter and the not
too long life-time of the plasma expected for RHIC conditions, the
propagated distance within deconfined matter is determined not only
by the geometrical size but mainly by the life-time $\tau_c$.
Therefore, $L$ depends sensitively on the initial thermodynamic state
of the deconfined matter.

To see some evidence of the specific QCD behavior of the energy loss
that includes the two regimes, i.e., the linear and quadratic $L$
dependences in Eq.~(1) governed in turn by the parameter
$L_{\rm crit} \propto \sqrt{E}$ in Eq.~(2),
we plot in Fig.~1 the dependence of the
average energy loss as a function of the initial transverse momentum
of the parent quark $p_\perp^0$.
For the central rapidity region we are considering one can
approximate $p_\perp^0 \approx E$.
The value of $p_\perp^0$ is most easily measured for photon tagged
jets since the high-energy photon does not suffer any noticeable
influence by the ambient medium.
(For estimates of the count rates
and so on cf.\ \cite{Wang}. 
By using the corresponding parts of PYTHIA \cite{PYTHIA} 
we have checked that intrinsic
parton transverse momenta before the hard scattering process
$g + q \to \gamma + q$ with momentum spread
$\sqrt{\langle k_\perp^2 \rangle} = 0.8$ GeV does not spoil the results.
In doing so we identified $p_\perp^0$ with $p_\perp^\gamma$.)
The results displayed in Fig.~1 refer to a transverse radius
$R_A = 7$ fm and an initial time of $\tau_i = 0.2$ fm/c.
In contrast to a simple
energy loss scenario with $\Delta E = \eta L$ with the widely used
constant $\eta = - 1$ GeV/fm,
one can observe indeed the two-regime behavior for the QCD based
energy loss according to Eqs.~(1, 2). Due to the above mentioned dependence
of $L$ on the life-time of the quark-gluon phase,
only in restricted ranges of initial temperatures and
initial jet momentum the inequality
$L > L_{\rm crit}$ is satisfied, i.e.,
a high enough initial temperature but not
too large jet momentum, so that the regime $\Delta E \propto L$
with increasing energy loss $\Delta E \propto \sqrt{E}$
for increasing $p_\perp^0$ becomes visible.
For large values of $p_\perp^0$ one enters ultimately the regime
$\Delta E \propto L^2$ which does not display any dependence on
$p_\perp^0$. Also for smaller initial temperatures and shorter
life-times $\tau_c$ and consequently smaller values of $L$,
one enters again the $p_\perp^0$ independent regime with
$\Delta E \propto L^2$. Notice that the energy loss, at larger values of
$p_\perp^0$, for the QCD behavior according to Eqs.~(1, 2) and the
linear model with $\Delta E = \eta L$ deliver similar results
for smaller values of $T_i$, while with increasing $T_i$ the energy loss
becomes larger in the former approach due to the $L^2$ dependence.

The initial temperature of the plasma, $T_i$  remains
as unknown parameter which is under intensive dispute now.
In particular within the saturation model \cite{EKRT}, where the
mechanism of the transverse energy production is determined by
gluon mini-jets \cite{Shuryak,Eskola1,Eskola2,PLB97},
the value of $T_i$ scales with $\sqrt{s}$, and
$T_i(\sqrt{s} = 200 \mbox{GeV}) \approx 0.6$ GeV is estimated.
As well known, the value of the initial temperature depends strongly
on the kinetic framework used to model the thermalization process
in the partonic system. Within the Boltzmann equation in relaxation
time approximation and imposed boost invariance \cite{Baym},
a lower initial
temperature of $T_i \approx 340$ MeV is found in \cite{GD}.
Under the reasonable assumption that the entropy is mainly
produced in the very early stage of the heavy-ion collision, where
$\tau \ll \tau_i$,
and the partonic matter
undergoes boost-invariant longitudinal expansion, the
combination $\tau_i T_i^3$ describes the entropy per unit rapidity interval
during both the pre-equilibrium and equilibrium stages and therefore
can be related to the multiplicity of secondary particles
measured experimentally.
Keeping this in mind we quantify the sensitivity of the QCD governed energy
loss of a fast quark on the initial thermodynamic state of matter
via its dependence on the parameter $N_i = \tau_i T_i^3$.

In Fig.~2 we plot the dependence of scaled energy loss
$(p_\perp^0 - \langle p_\perp \rangle )/N_i$ on the value $N_i$ for
various values
of the initial time $\tau_i$. For the initial energy of the fast
quark we take $E = p_\perp^0 = 50$ GeV
as appropriate for RHIC energies. Due to the
specific $L^2$ dependence (cf.\ Fig.~1) the scaled QCD energy loss increases
almost linearly with $N_i$. This is in contrast to the linear energy
loss model with $\Delta E = \eta L$ which is approximately independent of
the value of $N_i$. The above behavior of the QCD energy loss has
actually a qualitative character and is stable with respect to a
variation of the medium parameters in a wide region.
As already mentioned, to enter the $\Delta E \propto L^2$ regime
one has to satisfy first
the inequality $\tau_c < R_A$, where $R_A$ is the transverse radius of
the deconfined region, and second $\tau_c < L_{\rm crit}$,
since the average propagated distance to escape the
deconfined region is $\langle L \rangle \approx \tau_c$. Both these
inequalities reflect the not too long life-time of the deconfined stage
and a sufficiently high energy of the jet quark as realistic
for RHIC conditions.

To check the robustness of our results we also calculate the energy
loss by generating events with a probability
$\propto (1 - (r/R_A)^2)$, where $r$ is the radial coordinate
of the initial position of the hard quark. As seen in Fig.~3 the
increase of the scaled QCD energy loss becomes stronger in this case
while the linear energy loss model is still almost independent
of $N_i$.

\section{Angular dependence}

Since the bremsstrahlung gluons are mainly radiated in nearly 
the same direction
as the parent parton propagates and the energy loss is strongly dominated
by the gluon radiation, the experimental measurement of the energy
loss of the jet quark in QCD faces the problem of selecting the proper
angular size of the jet cone \cite{LS1,Dok}. 
This problem is studied in detail
for LHC conditions in \cite{LS}.
In our consideration we use the results of \cite{B3}
where the complete calculations of the angular dependence of the
energy loss are presented
(cf.\ also \cite{Urs} for a recent discussion of this topic).

The effect of the finite size of the jet cone for the parent parton's
energy loss can be characterized by a ''suppression'' factor
$R(\theta) = \Delta E(\theta) / \Delta E$, where $\Delta E(\theta)$
is the energy radiated away outside of an angular cone with
opening angle $\theta$, and $\Delta E$ is the total radiated energy.
As shown in \cite{B3}, the factor $R$ actually depends on the
combination $c(L) \theta$ with
$c(L) = \sqrt{\frac{N_c}{2C_f} \hat q ( \frac{L}{2})^3}$
and $C_f = \frac{N_c^2 - 1}{2 N_c}$.
While the explicit equations for $R(\theta)$ are somewhat involved,
the simple parameterization
$R(\theta) \approx \exp
\left(-a \sqrt{c(L) \theta} \right)$ with $a = 0.432$
results as appropriate fit displaying also the $L$ dependence.
As pointed out in the previous section, the averaged $L$ dependence
can be translated into a dependence on the initial thermodynamic
state parameters of the plasma.
To estimate the initial temperature dependence of the
factor $R$, which actually becomes $R(\theta, T_i)$,
we approximate here the propagation distance within
the Bjorken scenario by $L \approx \tau_i (T_i/T)^3$. Taking the
initial time $\tau_i = 0.2$ fm/c and using for the parameters
related to transport properties of the deconfined medium the same
values as described in the previous section we get
\begin{equation}
R(\theta, T_i) \approx \exp \left[
-5.306 \sqrt{ \theta } \left( \frac{T_i}{\mbox{GeV}} \right)^{9/4}
\right] .
\end{equation}
The corresponding temperature dependence for various angles
is displayed in fig.~4. The value of the characteristic
angle where $R \approx 1$, and consequently the parton energy
loss can be associated directly with the jet energy, decreases for increasing
initial temperature of the medium.
At the same time, for the initial temperatures expected at RHIC,
i.e., $T_i = 300 \cdots 550$ MeV, the characteristic angle is about
$0.01 \cdots 0.1$ rad which looks acceptable for an experimental
feasibility. That means one must measure the jet energy (or transverse
momentum at mid-rapidity) in a sufficiently small cone around
the jet axis to filter out the radiated energy.

The results of our calculations of the energy loss as a function
of the initial plasma temperature for various values of the angular cone
are displayed in fig.~5.
In accordance with the above analysis  for the temperature range expected
for RHIC one can notice only a tiny modification of the total
radiative energy loss if the jet cone
is restricted to $\theta \sim 0.01 \cdots 0.1$ rad. Basing on results of
\cite{LS} one can also conclude that within such a jet cone the radiative
energy loss is much greater than the collisional one and
can be related directly to the final jet energy.

\section{Conclusions and discussion}

In summary we have considered the dependence of the QCD radiative
energy loss of a fast parton on the initial thermodynamic state
of deconfined matter for conditions expected to be achieved
in central heavy-ion collisions at RHIC.
One of the appropriate ways to pin down the initial state is to vary
the beam energies in central collisions
which in turn should change the life-time of the
deconfined matter. If the life-time of the deconfined matter is less
than the time a fast parton needs to pass the typical geometrical size
of matter, the jet energy loss depends strongly on this life-time
due to the specific QCD governed $L^2$ dependence.
We show that such a QCD behavior causes an increase of the
jet energy loss with increasing entropy density of the system which
is encoded in the combination $\tau_i T_i^3$. For the widely used
value of $\tau_i = 0.2$ fm/c the proper range of initial temperatures
is 300 $\cdots$ 550 MeV in agreement with usual predictions for
RHIC. For a sufficiently high initial temperature of deconfined matter
one can also expect a remarkable dependence of the parton energy loss
on the initial parton energy which can be tagged by a hard photon:
Two different regimes, related to two different QCD radiation patterns
should become visible.

Our conclusions on the final jet energy loss are obviously valid if the QCD
energy loss  of the initial parton is much greater than the corresponding
loss in the hadronic medium and if the collisional energy loss is
negligible. In practice one needs to select a small enough
angular cone around the jet axis to measure only the energy
of the fast parent parton. 
For initial temperatures achieved at RHIC 
the opening angle should be not larger than 0.1 rad.

\subsection*{Acknowledgments}

Stimulating discussions with
R. Baier, X.N. Wang, and G.M. Zinovjev
are gratefully acknowledged.
O.P.P. thanks for the warm hospitality of the nuclear theory group
in the Research Center Rossendorf.
This work is supported in parts by BMBF grant 06DR921,
WTZ UKR-008-98 and STCU-015.

\begin{figure} 
\centerline{\psfig{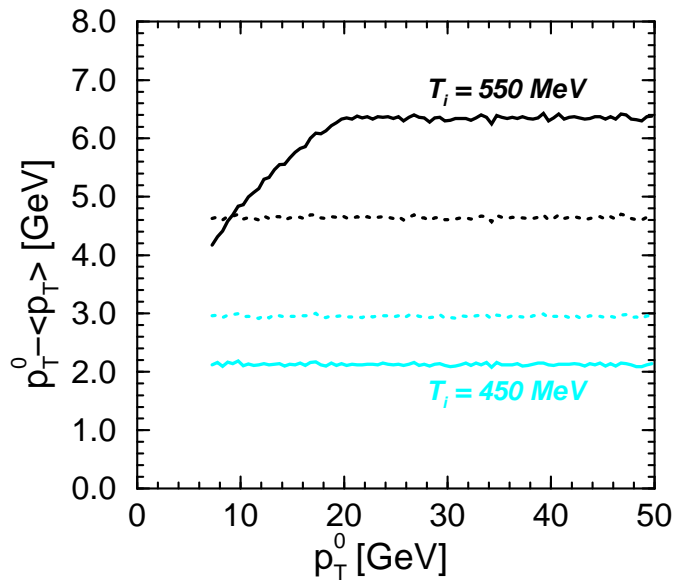}}
\caption{
The measure of the energy loss
$p_\perp^0 -  \langle p_\perp \rangle$
as a function of the initial parton momentum $p_\perp^0$
for the QCD energy loss according to Eqs.~(1, 2) (solid curves)
and for the linear energy loss model $\Delta E = \eta L$ with
$\eta = -1 $GeV/fm (dotted lines).
The black (gray) curves are for $T_i = 550$ (450) MeV.
}\end{figure}

\begin{figure} 
\centerline{\psfig{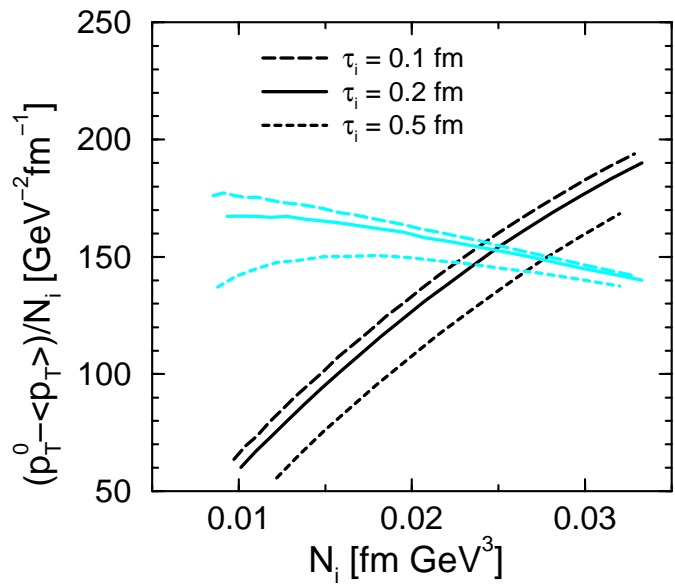}}

\caption{
The scaled energy loss as a function of $N_i = \tau_i T_i^3$
for various initial times $\tau_i$.
Black (gray) curves are for the QCD energy loss according to
Eqs.~(1, 2) (linear energy loss model $\Delta E = \eta L$ with
$\eta = -1 $GeV/fm).
}\end{figure}

\begin{figure} 
\centerline{\psfig{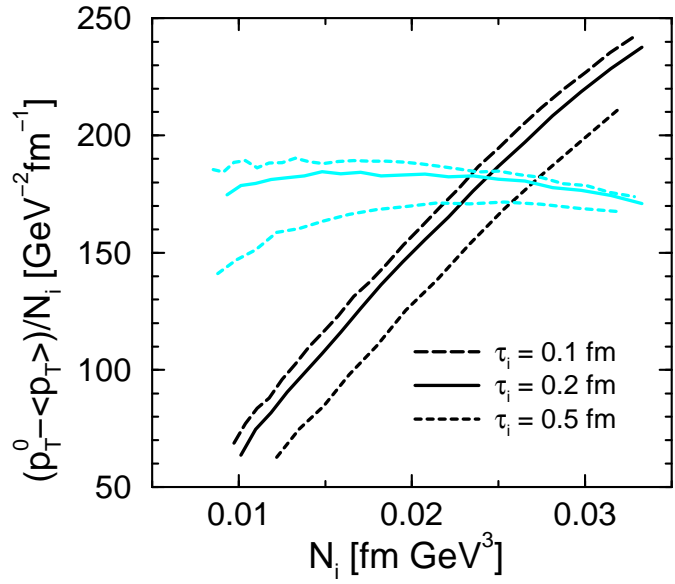}}

\caption{
The same as Fig.~2 but for another radial distribution
of initial positions as described in text.
}\end{figure}

\begin{figure} 
\centerline{\psfig{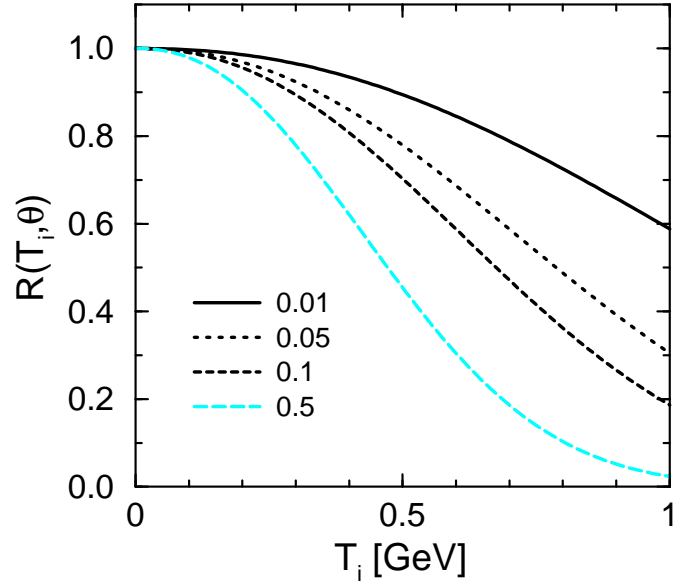}}

\caption{
The suppression factor $R(T_i, \theta)$ as a function of $T_i$
for various values of the cone angle $\theta$ (in units of rad).
}\end{figure}

\begin{figure} 
\centerline{\psfig{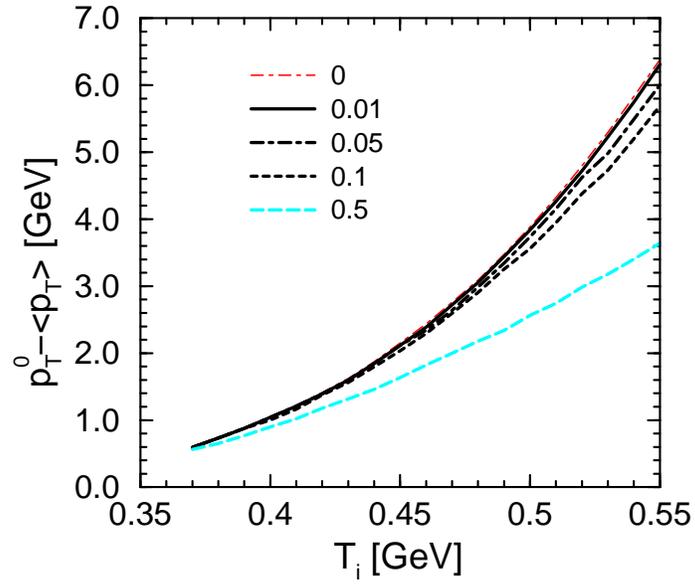}}

\caption{
The energy loss as a function of $T_i$ for various opening
angles of the jet cone (in units of rad).
}\end{figure}
\end{document}